\documentstyle[12pt,aaspp4,flushrt,psfig]{article}
\begin{document}

%

\title{ON THE ORIGIN OF THE IRON K LINE IN THE SPECTRUM OF THE
GALACTIC X-RAY BACKGROUND} 

\author{
Azita Valinia\altaffilmark{1,2}, Vincent Tatischeff\altaffilmark{3},
Keith Arnaud\altaffilmark{1,2},  
Ken Ebisawa\altaffilmark{1,4}, 
and Reuven Ramaty\altaffilmark{1}}  

\altaffiltext{1}{Laboratory for High Energy
Astrophysics, Code 662, NASA/Goddard
Space Flight Center, Greenbelt, MD 20771; valinia@milkyway.gsfc.nasa.gov}
\altaffiltext{2}{Department of Astronomy, University of Maryland,
College Park, MD 20742}
\altaffiltext{3}{Center de Spectrom\'etrie Nucl\'eaire et de Spectrom\'etrie de
Masse, IN2P3-CNRS, 91405 Orsay, France}
\altaffiltext{4}{USRA} 

\begin{center}
Accepted for Publication in the Astrophysical Journal
\end{center} 

\begin{abstract}
We propose a mechanism for the origin of the Galactic ridge X-ray
background that naturally explains the properties of the Fe K line,
specifically the detection of the centroid line energy below 6.7~keV
and the apparent broadness of the line.  Motivated by recent evidence
of nonthermal components in the spectrum of the Galactic
X-ray/$\gamma$-ray background, we consider a model that is a mixture
of thermal plasma components of perhaps supernova origin and
nonthermal emission from the interaction of low energy Cosmic ray
electrons (LECRe) with the interstellar medium.  The LECRe may be
accelerated in supernova explosions or by ambient interstellar plasma
turbulence.  Atomic collisions of fast electrons produce
characteristic nonthermal, narrow X-ray emission lines that can
explain the complex Galactic background spectrum.  Using the ASCA GIS
archival data from the Scutum arm region, we show that a
two-temperature thermal plasma model with $kT \sim 0.6$ and $\sim
2.8$~keV, plus a LECRe component models the data satisfactorily.  Our
analysis rules out a purely nonthermal origin for the emission.  It
also rules out a significant contribution from low energy Cosmic ray
ions, because their nonthermal X-ray production would be accompanied
by a nuclear $\gamma$-ray line diffuse emission exceeding the upper limits
obtained using OSSE, as well as by an excessive Galaxy-wide Be
production rate.  The proposed model naturally explains the observed
complex line features and removes the difficulties associated with
previous interpretations of the data which evoked a very hot thermal
component ($kT \sim 7$~keV).
\end{abstract} 
 
\keywords{galaxies: individual (Milky Way) --- ISM: structure --- X-rays: ISM ---
cosmic rays --- supernova remnants }

\newpage
%
 
\section{INTRODUCTION}

The origin of the Galactic X-ray background has remained a puzzle
since its discovery during the pioneering rocket experiments of Bleach
et al. (1972).  Since then, X-ray emission from the Galactic plane
and, in particular, the ridge (the narrow region around the Galactic
center covering from $\pm 60^\circ$ in longitude and $\pm 10^\circ$ in
latitude) has been studied with every major X-ray observatory.
Below 10~keV, the spectrum is rich in emission lines from Fe, Mg, Si
and S (e.g. Kaneda et al. 1997; hereafter K97). Above 10~keV, a power
law component is clearly detected (e.g. Yamasaki et al. 1997; Valinia
\& Marshall 1998) with an exponential cutoff near 40~keV
(Valinia, Kinzer \& Marshall 2000, hereafter VKM00).

Both diffuse and discrete sources have been proposed for the origin of
this emission. RS~CVn binaries and Cataclysmic Variables (CVs) have
been considered as significant contributors to the emission below
10~keV. However, Ottmann \& Schmitt (1992) have estimated that only
1\% of the iron K line detected in the unresolved Galactic background
spectrum is due to RS~CVns. Kaneda (1997) has shown that in order for
CVs to make up the diffuse emission, their space density would have
to be a factor of ~100 larger than previously estimated
(Patterson 1984). Recently, from the analysis of ROSAT PSPC data,
Tanaka, Miyaji \& Hasinger (1999) have shown that a class of hard
sources with $L_X \sim 10^{29-30} \, {\rm erg \, s^{-1}}$ and as
abundant as stellar coronal sources is required to account for the
unresolved ridge emission.  No such class of sources has been
identified.

The discovery using {\it Tenma} of the iron K line in the ridge
spectrum motivated the idea that the emission below 10~keV is due to a
diffuse, hot, optically-thin plasma of temperature 5--15~keV (Koyama
et al. 1986).  Later, high resolution measurements of the ridge
spectrum using ASCA revealed a complex emission line structure. The
ASCA spectrum cannot be satisfactorily modeled with a single thermal
component in ionization equilibrium (K97).  Instead, K97 modeled the
emission by two non-equilibrium ionization components with
temperatures of $\sim 0.8$ and 7~keV. While this model provides a good
fit to the spectrum, it is difficult to explain the production and
confinement of such hot gas uniformly permeating the interstellar
medium (ISM). Supernova remnant shock temperatures reach only about
3~keV. Also, the 7~keV temperature exceeds the gravitational potential
of the disk by at least one order of magnitude (Townes
1989). Therefore, according to this model, the hot gas must be
constantly and uniformly replenished in the Galaxy.  To overcome this
difficulty, Tanuma et al. (1999) consider heating and confinement of
the plasma in the Galactic plane via magnetic reconnection. They
numerically investigate how magnetic reconnection triggered by
supernova explosions heats the plasma which is consequently confined
in magnetic islands.  They conclude that for this mechanism to work,
local magnetic fields of $\sim 30 \, \mu$G are necessary. This model
implies that the hot plasma must be very localized.

Another problem with the purely thermal scenario is that the
two-temperature model can not account for the data above 10~keV, which
require at least one additional component at higher
energies. Simultaneous RXTE and OSSE observations up to 1~MeV show the
presence of an additional component in the spectrum that can be
characterized as an exponentially cut-off power law (VKM00). This
component dominates in the 10-400~keV band and contributes
approximately 35\% of the total 3-10~keV emission. VKM00 estimated
that up to $\sim 50\%$ of the emission at 60~keV is due to discrete
sources, with the rest remaining unresolved. These results
imply that there is a nonthermal contribution to the Galactic ridge
emission below 10 keV. This paper explores some of the consequences.

The unresolved hard X-ray/soft $\gamma$-ray diffuse component detected
by RXTE and OSSE can be produced either by bremsstrahlung from low
energy Cosmic ray electrons ($< 1$~MeV; LECRe) or inverse Compton
scattering off ultra-relativistic ($\sim$0.1-10 GeV) electrons (e.g.
Skibo \& Ramaty 1993). However, Skibo, Ramaty \& Purcell (1996) have
shown that the emission up to 100~keV is more likely of electron
bremsstrahlung origin, because a large population of
ultra-relativistic electrons would also produce a significant Galactic
synchrotron radio emission at $\sim$100 Mhz and this is not
observed. In addition to the bremsstrahlung continuum, LECRe produce
nonthermal narrow X-ray emission lines below 10~keV via K-shell
vacancy creation in ambient heavy atoms.  Low energy Cosmic ray ions
(LECRi) also produce characteristic broad and narrow X-ray emission
lines (Tatischeff, Ramaty \& Kozlovsky 1998 and references therein,
Dogiel et al. 1998). Their contribution to the Galactic ridge diffuse
emission has been considered by Tatischeff, Ramaty \& Valinia (1999)
and Tanaka, Miyaji \& Hasinger (1999). However, we show in \S 4 that
LECRi can not account for the Galactic X-ray background because their
nonthermal X-ray production would be accompanied by a nuclear
$\gamma$-ray line diffuse emission exceeding the upper limits obtained
using OSSE, as well as by an unreasonably high Galaxy-wide Be
production rate. For these reasons, we focus on the role of low energy
Cosmic ray electrons in nonthermal X-ray production below 10~keV.

In this paper, we report on the analysis of $\sim 111$~ks archival
ASCA data from the Scutum arm region on the Galactic plane.  Instead
of applying non-equilibrium ionization models, we show that an
alternative model, namely nonthermal X-rays from the interaction of
LECRe with the ISM plus two thermal plasma components (of $\sim
2.8$~keV and 0.6~keV) produces an equally good fit to the
spectrum. According to our best fit model, the thermal and nonthermal
components contribute
81\% and 11\%, respectively, of the $0.6-9$~keV emission,
with the remaining emission contributed by the Cosmic X-ray background. 
(In the 4-9~keV band, the thermal and nonthermal components contribute
38\% and 36\%, respectively.)
In \S 2 we investigate the X-ray line and continuum emission from the 
interaction of LECRe with the ISM. In \S 3 we present 
the data and our analysis and in \S 4 we discuss the implications 
of our results. 

\section{X-RAYS FROM LECRe INTERACTION WITH THE ISM}

In the LECRe scenario, nonthermal 
X-ray line emission results from the filling of 
inner-shell vacancies produced by the fast electrons in the ambient atoms. 
We assumed the X-ray production region to be 
neutral and of solar composition (Anders \& Grevesse 1989) and considered 
the K$\alpha$ and K$\beta$ lines from ambient C, N, O, Ne, Mg, 
Si, S, Ar, S and Fe. The transition 
energies are given in Tatischeff et al. (1998, Table~6). The X-ray line 
production cross section can be written as 
\begin{eqnarray} 
\sigma_{Line}=\sigma_{I}wk, 
\end{eqnarray}
where $\sigma_{I}$ is the cross section for the collisional
ionization leading to the K-shell vacancy, $w$ is the
fluorescence yield for the K shell (Krause 1979) and $k$ is the
relative line intensity among the possible transitions which
can fill the inner-shell vacancy (Salem, Panossian, \& Krause
1974). For the K-shell ionization cross section, we used the
semiempirical formula of Quarles (1976), which is in good agreement
with the data compiled in Long et al. (1990). Since the intrinsic 
width of the X-ray lines is smaller than the ASCA energy resolution, 
the determination of the 
exact line profiles is beyond the scope of this paper. 
For simplicity, we adopted 
the same width for all the lines, $\Delta E_X$=10 eV, which is the
approximate width of the 6.40 keV Fe line in a cold ambient medium (Dogiel
et al. 1998). 

X-ray continuum emission is essentially due to bremsstrahlung from
the fast primary electrons. We estimated bremsstrahlung from
knock-on secondary electrons to be of minor importance above 0.5 keV. We 
considered primary electron bremsstrahlung in ambient H and He and
calculated the corresponding differential cross sections from 
equation (3BN) in Koch \& Motz (1959).

The differential X-ray production rate can be written as
\begin{eqnarray} 
{dQ \over dE_X}= \sum_{j} n_j \int_{0}^{\infty} J_e(E) v(E)
{d\sigma_{j} \over dE_X} (E_X,E) dE~, 
\end{eqnarray}
where $n_j$ is the density of the ambient isotope $j$, $J_e(E)$ is the
differential equilibrium electron number as a function of the electron
kinetic energy $E$, $v(E)$ is the electron velocity and
$d\sigma_{j} / dE_X$ is the differential X-ray production cross section
for electron interactions with particles of type $j$. We perform
calculations with an electron equilibrium spectrum of the form
\begin{eqnarray} 
J_e(E) \propto E^S e^{-E/E_0}~. 
\end{eqnarray}

Figure 1 shows the X-ray emission produced by a LECRe population
with $S$=0.3 and $E_0$=90 keV, which provides a good
fit to the RXTE and OSSE Scutum arm data (see \S 3). The calculations are
normalized to a total kinetic energy in the fast electrons
$E_{tot}$=$\int_{0}^{\infty}EJ_e(E)dE$=1 erg and an average ambient H
density of $1 \, {\rm atom \, cm^{-3}}$. The spectral index $S$=0.3 could
be due to the injection of electrons of energies $\gtrsim E_0$
into the neutral phase of the ISM. Indeed, the equilibrium electron spectrum 
would then satisfy
\begin{eqnarray} 
J_e(E<E_0) = {1 \over dE/dt} \int_{E_0}^{\infty} {dN_e \over dt} (E') dE'
\propto {1 \over dE/dt}~,
\end{eqnarray}
where $dN_e/dt$ is the electron source spectrum (it is assumed that
$dN_e/dt(E)=0$ for $E<E_0$) and $dE/dt$ is the electron energy loss rate
in a neutral ambient medium, which is roughly proportional to $E^{-0.3}$
below $\sim$100 keV (Berger \& Seltzer 1982). As shown in Figure~1, the 
continuum emission is accompanied by strong X-ray lines in the ASCA 
energy range. 

\begin{figure}[tbh]
\centerline{
{\hfil
\psfig{figure=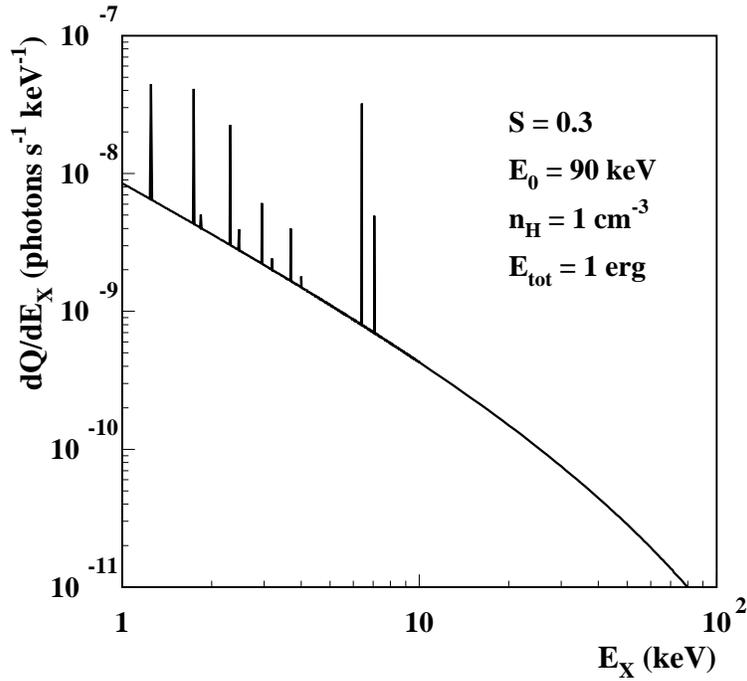,height=4.0truein,angle=0.0}
\hfil}
}

\caption
{\footnotesize Nonthermal X-ray emission produced by LECRe with $S$=0.3 and
$E_0$=90 keV (eq.~[3]) interacting in a neutral ambient medium of solar
composition. The differential X-ray production rate is normalized to an
average ambient H density of 1 atom cm$^{-3}$ and a total electron kinetic
energy of 1 erg.} 
\end{figure} 

\section{OBSERVATIONS AND ANALYSIS} 

We used archival ASCA GIS data of the on-plane (i.e. $b \sim 0^\circ$)
Scutum arm region from Galactic longitude $26^\circ\!.5 < l <
30^\circ\!.0$.  Overall, 10 pointings with a total good time exposure
of $\sim 111$~ks were used.  These observations were performed from
1993 to 1997. For spectral analysis, data from the central region of
the GIS with radius less than $22^\prime$ was used, as this is the default
cut-off radius in the ASCA Standard Processing Products.  Within this
radius, spectral resolution is known to be uniform, and the long term
gain change and positional gain variation is calibrated within $\pm 1
\%$ (Makishima et al. 1996).

The average count rate in the 0.5-10 keV band after removing point
sources was $0.526 \, {\rm cts \, s^{-1}}$ for GIS2 and $0.559 \, {\rm
cts \, s^{-1}}$ for GIS3.  We were not able to make significant use of
SIS data since fewer photons were collected by those instruments (the
field of view for the SIS being much smaller than that for the GIS)
and furthermore the degradation of the SIS response in later
observations reduced the utility of that instrument.

Since the emission region is larger than the GIS field-of-view,
``stray light effects'' have to be taken into account. For the purpose
of this analysis, an XRT response (i.e.  ancillary response file or
ARF) has been constructed by taking into account the stray-light
effects up to $1^\circ\!.5$ from the XRT axis.  The procedure is as
follows.  We randomly generated one-million X-ray photons within
$1^\circ\!.5$ of the optical axis, and using ray-tracing calculated
the number of photons which reached the detector plane. We repeated
this procedure for 300 monochromatic energies from 0.1 to 12~keV.
From these ray-traced images, taking account of the GIS detector
response, we calculated the ARF for the diffuse emission corresponding
to the detector region used (which excluded any point sources). For
each energy bin,
\begin{equation}
{\rm ARF (cm^2 str)} \propto  {\rm S (cm^2)} \times  \Omega {\rm (str)}
 \times {N \over 10^6},
\end{equation}
where $S$ is the geometrical opening area of the ASCA X-ray telescope
($826.64 \, {\rm cm^2}$), $\Omega$ is the solid angle of the assumed
spatial extent ($2.153 \times 10^{-3}$~str), and $N$ is the number of
photons to have reached the detector region concerned.  This is
basically the same method described in K97.

We also calculated the diffuse ARF for a more realistic distribution
of the emission along the Galactic plane.  We assumed that the spatial
distribution of emission was Gaussian in latitude with a $4^\circ$
FWHM (as measured by Valinia \& Marshall 1998) and constant in
longitude.  We found an insignificant difference in the best fit model
parameters with the two different ARFs. In what follows we show
results obtained with the ARF made assuming uniform emission from a
circle of radius $1^\circ \! .5$.

Since the Cosmic X-ray background (CXB) is absorbed by the Galactic
plane at energies below $\sim 5$~keV, we subtracted only the
instrumental background from the data and included the CXB as a fixed
component in the model. Archival night earth observations were used
to determine the instrumental background. From the analysis of the
blank sky GIS2 and GIS3 archival data, we found that the CXB was best
fitted with a power law function $k(E/1~{\rm keV})^{-\Gamma}$. The
best fit parameters were $\Gamma \sim 1.4$ and $k = $ $0.0177$ and
$0.0192$ ${\rm photons \, cm^{-2} \, s^{-1} \, keV^{-1}}$ at 1~keV for
GIS2 and GIS3, respectively. In all the fits reported below
the CXB component was absorbed by the best-fit Galactic
hydrogen column.  We simultaneously fit the GIS2 and GIS3
spectra.  Because the emission is faint, we found that many channels
were short of the $\sim 20$ counts minimum required to yield
meaningful $\chi^2$ statistics. Therefore, before fitting the data, we
binned channels 0-511 by a factor of 2 and channels 512-1023 by a
factor of 8. This resulted in at least 20 counts in each bin.

Since the complex shape of the Fe K line is the prime focus of this
paper, we first analyzed  the 4-9~keV data (the channels above 9~keV have
insufficient counts to be worth including in the fit). At
these energies, the spectrum is little affected by the absorption due
to the Galactic hydrogen column density.  Our first attempt was to
model the data using a single thermal plasma component (the mekal
model; Mewe, Gronenschild, \& van den Oord 1985; Mewe, Lemen, \& van
den Oord 1986; Kaastra 1992) absorbed by the Galactic plane.
We derived a best fit temperature of $kT = 4.7{\pm 1.0}$ keV.
The result of this fit is shown in Figure 2a. The fit around the Fe K
line is very poor. Therefore, we modeled the data using a thermal
bremsstrahlung plus a Gaussian line component, as was also done by
K97. The result of this with a best fit temperature parameter of $kT =
7.4^{+2.8}_{-2.3}$ keV is shown in Figure 2b. Notice that the quality
of the fit around the Fe K line has dramatically improved but the 
width ($187{\pm 59}$~eV) suggests that the line is
intrinsically broad.  Also, the peak line energy is detected at
$6.627\pm0.035$~keV, a lower energy than that expected from a plasma
in ionization equilibrium.

\begin{figure}[tbh]
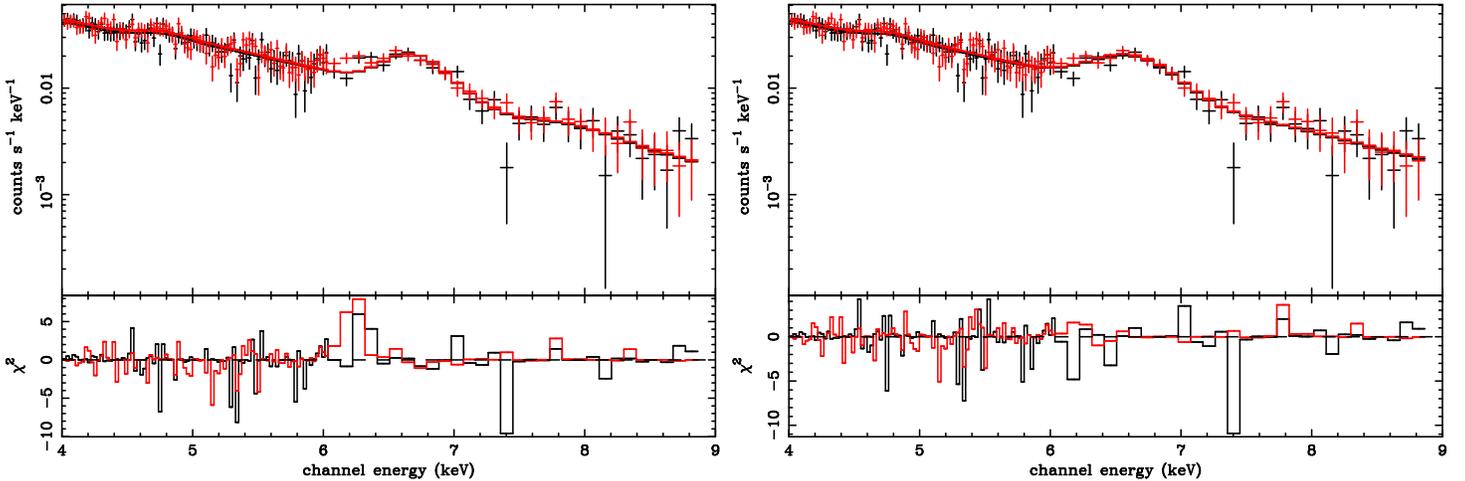

\centerline{
{\hfil \hfil
\psfig{figure=figure2a_color.ps,height=2.5truein,angle=270.0}
\psfig{figure=figure2b_color.ps,height=2.5truein,angle=270.0}
\hfil \hfil} }
\caption{\footnotesize 
(a) ASCA GIS2 and GIS3 data fitted to an absorbed mekal model (left
panel).
(b) Same data as in (a) fitted to an absorbed bremsstrahlung
plus Gaussian model (right panel).}
\end{figure}

Next, we modeled the data with the nonthermal LECRe
component. However, since the parameters of this component are
directly related to the emission above 10~keV, we {\it first} used the
RXTE and OSSE Scutum arm measurements of VKM00 to constrain the shape
of the electron equilibrium spectrum (i.e. $S$ and $E_0$ in eq.~3).
VKM00 modeled the 3-400 keV spectrum with an absorbed thermal model
plus an exponentially cutoff power law. We replaced the exponentially
cutoff power law with the LECRe model and found that any spectral
index $S>0$ provided equally satisfactory fits to the data. We fixed
the spectral index at $S=0.3$, consistent with the scenario that
accelerated electrons of energies $\gtrsim E_0$ are injected into a
neutral medium (see \S 2).  For the exponential turn-over energy, we
found a best fit $E_0 \sim 90$~keV. The best fit result is obtained
when it is assumed that the broad band diffuse emission has a spatial Gaussian
distribution of FWHM $\lesssim 0^\circ\!.5$ which is consistent 
with the interpretation that the hard X-ray/soft $\gamma$-ray emission
is confined to a $\lesssim 1^\circ$ thin disk centered roughly on the
Galactic midplane. Indeed, from the RXTE survey of the Galactic plane,
Valinia \& Marshall (1998) find two spatial components for the diffuse
emission: a thin disk of full width $\lesssim 0^\circ \!.5$ centered
roughly on the Galactic midplane and a broad component of Gaussian
distribution with FWHM of about $ 4^\circ$.  Moreover, this is consistent
with the observation that the Fe K line
is either extremely weak or not detected at all in ASCA measurements
of the Galactic plane diffuse emission at high latitudes, $l >
|1^\circ| $ (K97).  K97 also found that the scale height of the hard
X-ray component was smaller than that of the soft component.

Using the parameters for the LECRe model obtained from the high energy
data ($S=0.3, \, E_0=90$~keV), we fit the 4-9 keV ASCA data with an
absorbed LECRe model.  However, this model did not produce a
satisfactory fit to the data. In particular, the quality of the fit
around the Fe K line was very poor.  We then modeled the data with an
absorbed LECRe plus a thermal plasma component. This provides a
satisfactory fit to the entire energy range, including the Fe K line.
This line appears to be broadened and shifted to 6.6 keV because it is 
a combination of the 6.4 keV LECRe line and the 6.7 keV from He-like
Fe. Because the thermal plasma component has a temperature of 
$\sim 3$~keV there is a negligible contribution from the 6.97 keV line
of H-like Fe. The parameters of this fit are shown in the top panel 
of Table~1. 

Extending this model down to 0.6~keV, we found that it provided a good
fit to the data only in the 2-9~keV band.  Below 2~keV, the data
significantly exceeded the model, requiring an additional low
temperature component. Adding a second thermal plasma component to the
above model and fitting for the best parameters yielded a good fit to
the overall spectrum. The parameters of this model are given in the
bottom panel of Table~1.  These results suggest that there is a nearby
low-temperature thermal component ($kT \sim 0.6$~keV) which accounts
for the soft X-ray emission of the Galactic background (as implied by
the low value of the hydrogen column density). The hard X-ray
component (i.e. $E > 2$~keV) is produced from superposition of a
thermal component ($kT \sim 2.8$~keV gas) and a nonthermal LECRe
component, absorbed through a larger hydrogen column.  Figure~3a shows
the data and folded model. Figure~3b shows the unfolded spectrum and
model components.

\begin{figure}[tbh]
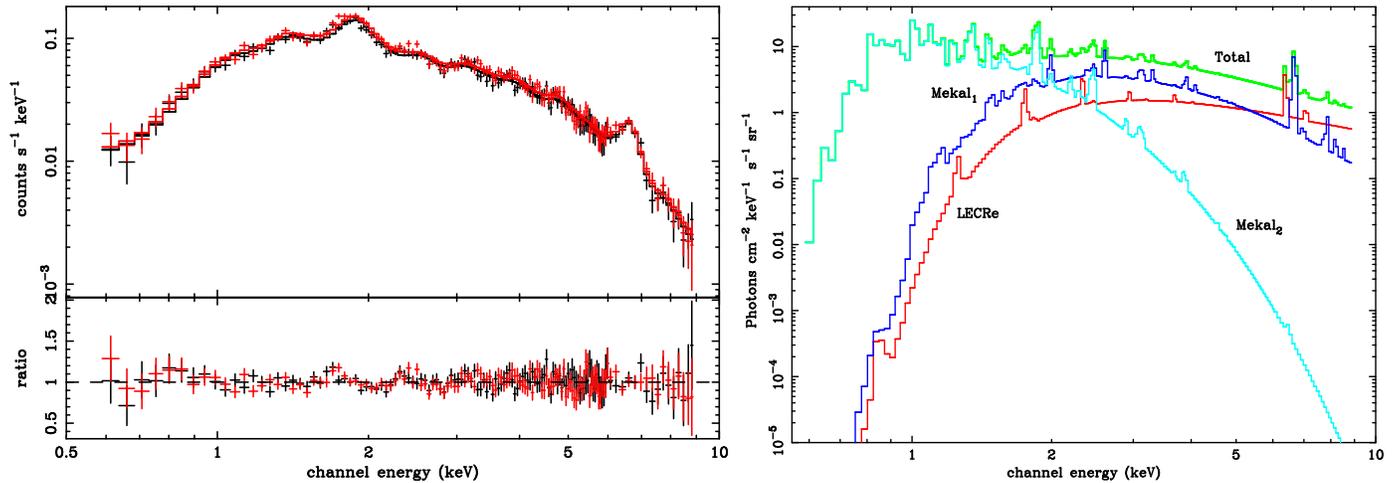

\centerline{
{\hfil \hfil
\psfig{figure=figure3a_color.ps,height=2.5truein,angle=270.0}
\psfig{figure=figure3b_color.ps,height=2.5truein,angle=270.0}
\hfil \hfil} }
\caption{\footnotesize (a) Data and folded model in our favorite emission scenario summarized in
Table~1 are shown.
The lower window shows the ratio of data divided by the folded model (left
panel).
(b) Unfolded spectrum and model components.
The CXB is included in the model
but not shown for clarity of the plot (right panel).
}
\end{figure} 

\section{DISCUSSION}

The composite thermal plus LECRe model in Table~1 provides the best
fit to the ASCA data. In this scenario, most of the emission below
2~keV is due to a thermal plasma component at $kT \sim 0.6$~keV.  The
hard X-ray emission above 2~keV is due to a combination of a hotter
plasma component at $kT \sim 2.8$~keV, which may have been shock
heated in supernova explosions, and a LECRe component which may also
be related to supernova explosions, and which would account for part
of the RXTE and OSSE data above 10 keV.  We have assumed that this
component is limited to the central $1^\circ$ of the Galactic plane.
This model produces a very good fit to the overall spectrum and, in
particular, to the Fe K line complex. This is because the apparent
broadening and shift in the mean energy of the Fe K line is produced
by a combination of the thermal lines from de-excitation in He-like Fe
and the nonthermal lines, primarily at 6.4~keV, due to K-shell
vacancy production in ambient neutral Fe.

The ASCA SIS should be able to resolve the nonthermal 6.4~keV line
from the thermal lines at $\sim 6.7$~keV. However, for the Scutum arm
region the signal-to-noise is not sufficient to draw any conclusions.
K97 have summed ASCA SIS data over a larger latitude range in order to
have enough counts to examine the shape of the Fe K line. They found
clear evidence for photons with energies below that of the He-like
line at $\sim 6.7$~keV and no evidence of emission at 6.4~keV. However, 
since the LECRe are likely confined close to the plane we expect any 
6.4~keV line to be diluted in the K97 data. The photons with energies
below that from the He-like line are from lower-ionization stages of
Fe and are not predicted by thermal emission from a collisional
plasma. As K97 correctly pointed out, this is good evidence for plasma
that is out of ionization equilibrium. We differ from K97 in not
requiring as high a continuum temperature so the plasma can be closer
to equilibrium.

The RXTE and OSSE Scutum arm data above 10 keV are best modeled by a hard
electron equilibrium spectrum ($S>0$) which could result from the injection
of electrons accelerated in a low-density ionized medium into a denser
neutral medium in which they would produce most of the nonthermal X-rays. 
Assuming an average H density of $1 \, {\rm cm^{-3}}$ in the X-ray
production region, we calculated that a total electron kinetic energy of
$2.9 \times 10^{53} \, {\rm erg}$ is needed to produce the 0.6-9 keV
luminosity in the LECRe component of $2.5 \times 10^{37} \, {\rm erg \,
s^{-1}}$ (Table~1). Dividing this suprathermal electron energy by the
estimated volume of the emission ($\sim 10^{66} \,{\rm cm^{3}}$) leads to
an energy density of $\sim 0.2 \, {\rm eV \, cm^{-3}}$, comparable to the
Cosmic ray ion energy density of $\sim 1 \, {\rm eV \, cm^{-3}}$ in the
Galaxy (e.g. Wdowczyk \& Wolfendale 1989). Therefore, on energetic grounds
the LECRe scenario is plausible. The LECRe may be accelerated in
supernovae remnants (e.g. Yamasaki et al. 1997) or by ambient interstellar
plasma turbulence (Schlickeiser 1997).

\begin{figure}[tbh]
\centerline{
{\hfil \hfil
\psfig{figure=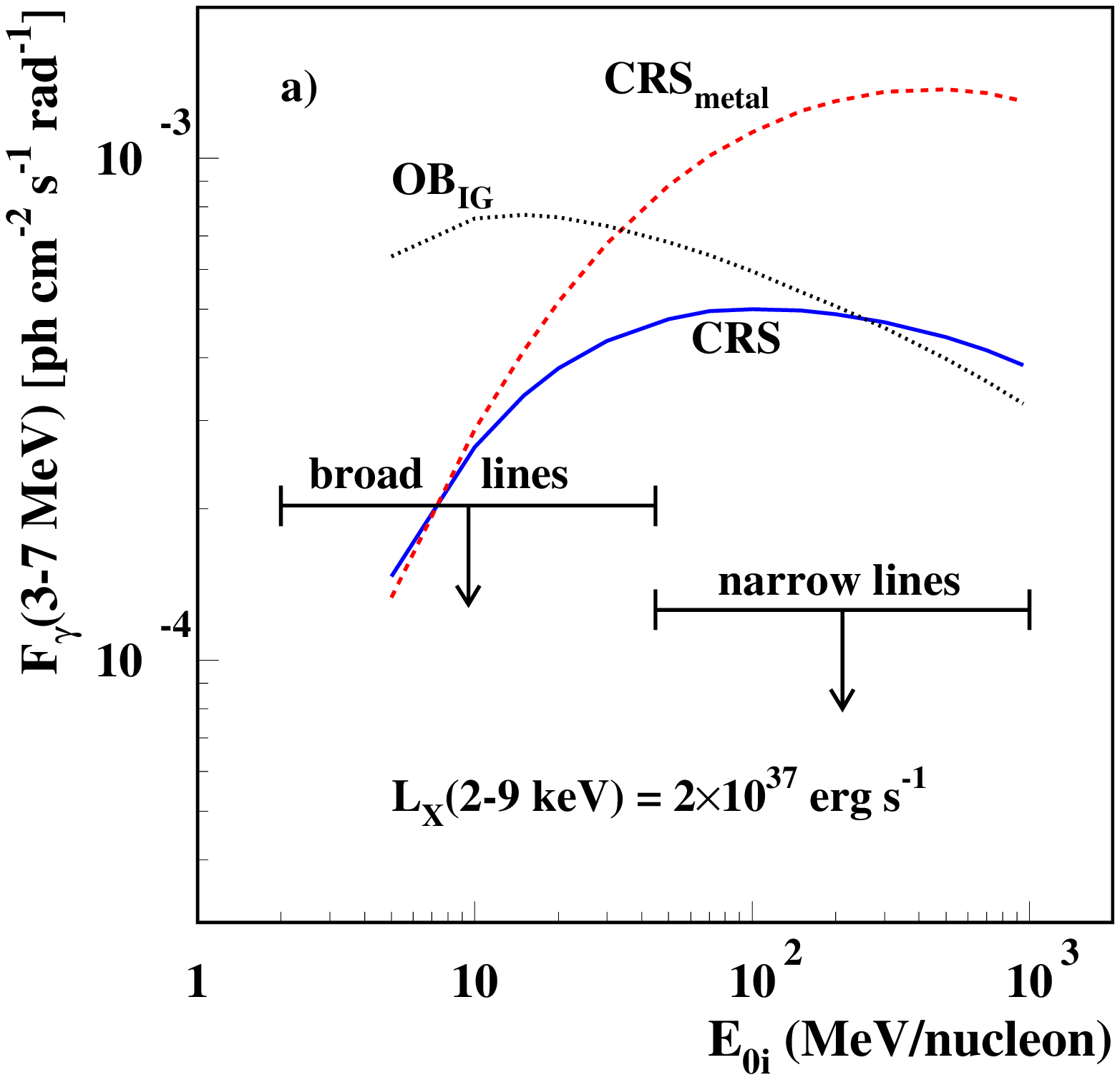,height=3.4truein,angle=0.0}
\psfig{figure=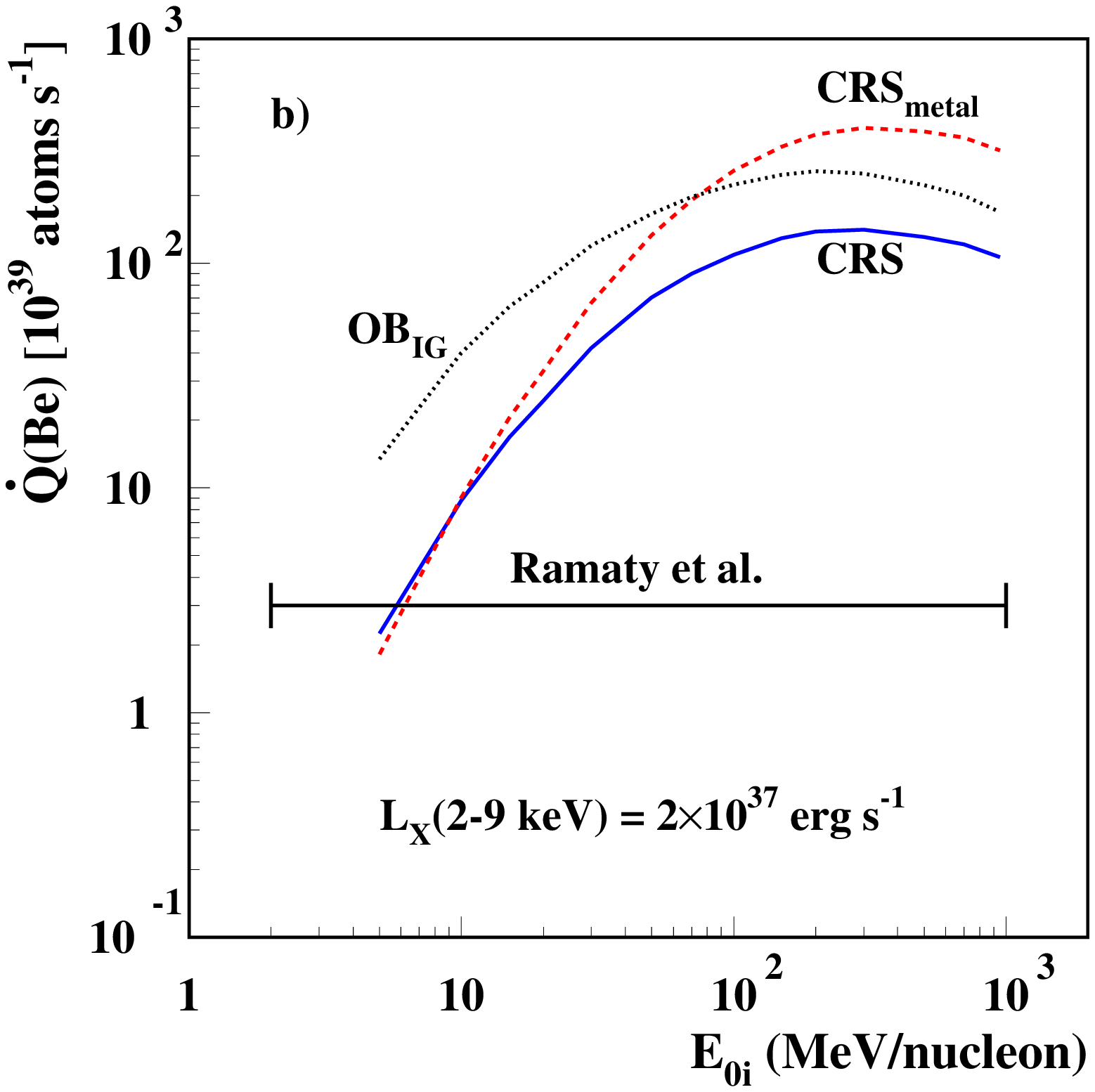,height=3.4truein,angle=0.0}
\hfil \hfil} }
\caption{\footnotesize 
(a) Calculated 3-7 MeV nuclear gamma-ray line fluxes from the central
radian of the Galaxy (left panel), and (b) predicted Galaxy-wide Be 
production rates,
as a function of $E_{0i}$ (eq.~[6]) (right panel). 
The calculations are normalized to a
nonthermal X-ray production by the fast ions of
$2 \times 10^{37} \, {\rm erg \, s^{-1}}$ in the 2-9 keV energy range. Also
shown in panels (a) and (b) are the OSSE upper limits for both broad and narrow
line emissions (Harris et al. 1997) and the current epoch Be
production rate recommended by Ramaty et al. (1997) and Ramaty et
Lingenfelter (1998), respectively.}

\end{figure}

We now show that low energy Cosmic ray ions (LECRi) can not contribute
significantly to the hard X-ray emission. LECRi which produce
nonthermal X-rays via atomic interactions also produce gamma-ray
emission and the light elements Li, Be and B via nuclear reactions
(Ramaty et al. 1997 and references therein). The calculated 3-7 MeV
nuclear gamma-ray line fluxes, and Galaxy-wide Be production rates
that would accompany the X-ray emission are shown in Figure ~4a and
4b, respectively. For these calculations, we used the same interaction
model as Ramaty et al. (1997) and Tatischeff et al. (1998) and an
accelerated ion source spectrum also used by these authors,
\begin{eqnarray} 
{dN_i \over dt} (E_i) \propto E_i^{-1.5} e^{-E_i/E_{0i}}~. 
\end{eqnarray}
We employed three different compositions for the LECRi: CRS and
CRS$_{\rm metal}$ (Ramaty et al. 1997) and OB$_{\rm IG}$ (Parizot,
Cass\'e \& Vangioni-Flam 1997).  CRS is the composition of the current
epoch Galactic cosmic-ray sources; CRS$_{\rm metal}$ is identical to
CRS, but without protons and $\alpha$ particles; OB$_{\rm IG}$ is the
calculated average composition of the stellar winds from OB
associations in the inner Galaxy. The calculations are normalized to a
nonthermal X-ray production by the fast ions of $2 \times 10^{37} \,
{\rm erg \, s^{-1}}$ in the 2-9 keV energy range, which is the
luminosity of the LECRe component in our best fit model (Table~1). In
figure~4a, the predicted gamma-ray line fluxes are compared with the
upper limits obtained with OSSE for both broad and narrow line
emissions from the central radian of the Galaxy (Harris et
al. 1997). The CRS$_{\rm metal}$ composition produces only broad
gamma-ray lines, whereas the CRS and OB$_{\rm IG}$ compositions also
produce intense narrow lines, from the interactions of accelerated
protons and $\alpha$ particles with ambient $^{12}$C and
$^{16}$O. Thus, only the CRS$_{\rm metal}$ composition with
$E_{0i}\lesssim$7 MeV/nucleon does not violate the OSSE upper limits
(Fig.~4a). Furthermore, it is shown in Figure~4b that the predicted Be
production rates significantly exceed the nominal value obtained from
the observation of the linear correlation between Be and Fe for
[Fe/H]$<$-1, $\dot{Q}({\rm Be})$=$3 \times 10^{39}$ Be-atoms s$^{-1}$
(Ramaty et al. 1997, Ramaty \& Lingenfelter 1998), with the exception
of very low energy ions of CRS$_{\rm metal}$ or CRS
composition. Moreover, nonthermal X-ray production by such very low
energy ions is very inefficient. We find that LECRi with
$E_{0i}\lesssim7$ MeV/nucleon should deposit more than $3 \times
10^{42} \, {\rm erg \, s^{-1}}$ in the ISM to produce the 2-9 keV
luminosity of $2 \times 10^{37} \, {\rm erg \, s^{-1}}$, exceeding by
more than a factor of 2 the total power delivered by Galactic
supernovae, $1.5 \times 10^{42} \, {\rm erg \, s^{-1}}$ (Ramaty \&
Lingenfelter 1998).  We thus conclude that LECRi do not significantly
contribute to the hard X-ray emission below 10~keV. Using different
ion source spectra, Pohl (1998) and Tatischeff, Ramaty \& Valinia
(1999) have shown that LECRi can not account for the X-ray data above
10 keV either.

Although the resolution of the ASCA data does not allow us to
distinguish between purely non-equilibrium ionization and LECRe plus
thermal models, we contend that the non-equilibrium ionization models
are generally disfavored. This is because the purely thermal models
require a very hot gas ($kT \sim 7$~keV) that is uniformly produced
and confined in the Galactic disk (a very contrived scenario). On the
other hand, the presence of a nonthermal component for the emission is
demonstrated by hard X-ray/soft $\gamma$-ray observations of the
Galactic background (e.g. Yamasaki et al.  1997; Valinia \& Marshall
1998; VKM00; Boggs et al. 1999).  The LECRe model along with a shock
heated plasma component (both of which may have been created in
supernovae explosions) provide a plausible explanation for the diffuse
Galactic background emission.  Observations performed with XMM are
expected to be able to distinguish between these scenarios because of
its high throughput and CCD spectral resolution.

\acknowledgements We thank Keith C. Gendreau for his help in preparing
the diffuse ASCA ARF during the early phase of this work.  This
research has taken advantage of HEASARC archival data base.

%

\newpage

{\footnotesize
\begin{deluxetable}{lll}
\tablecolumns{3}
\tablecaption{Models and Best Fit Parameters to the Galactic X-ray Background
Spectrum}
\tablehead{  \multicolumn{3}{c}{$4-9$ keV; $N_H({\rm Mekal+
LECRe+CXB\tablenotemark{a}}\,\,)$} \nl \cline{1-3}
Model &  Model Parameter & Value  }
\startdata
  &  $N_H$  &  $2.7^{+4.5}_{-2.7}$      \nl \\
  & $kT$(keV)   &   $2.8^{+1.2}_{-1.1}$     \nl
Mekal   & Abundances & $1.3^{+1.7}_{-0.5}$   \nl
  &  ${\rm Flux(\%)}$\tablenotemark{b}   &  38      \nl \\
  &  $s$   &  $0.3$  \nl
LECRe   &  $E_0$(keV)  &    90   \nl
  &  ${\rm Flux(\%)}$\tablenotemark{b}  &  36    \nl \\

  &  Total Flux\tablenotemark{c} & $1.39\times 10^{-7}$     \nl \cline{1-3}
 &   $\chi^2/\nu$  & 194.8/225       \nl  \cline{1-3}

\multicolumn{3}{c}{$0.6-9$ keV; $N_{H1}({\rm Mekal_1+LECRe+CXB\tablenotemark{a}
}\,\,)+N_{H2}(
{\rm Mekal_2})$} \nl \cline{1-3}
  &   $N_{H1}$  &  $3.6^{+0.5}_{-0.7}$  \nl \\
  &  $kT$(keV)  &  $2.8{\pm0.4}$  \nl
 ${\rm Mekal_1}$  &  Abundances &  $1.2^{+0.4}_{-0.3}$  \nl
  &   ${\rm Flux(\%)}$\tablenotemark{b}   &  24 \nl \\

  &  $s$   &  $0.3$  \nl
LECRe  &   $E_0$(keV)  &    90   \nl
  &  ${\rm Flux(\%)}$\tablenotemark{b}  &  11    \nl \\

  &  $N_{H2}$  & $1.4^{\pm0.1}$   \nl \\
  &  $kT$(keV) &   $0.56^{+0.02}_{-0.08}$ \nl
${\rm Mekal_2}$   &  Abundances &   $0.5^{+0.5}_{-0.1}$   \nl
  &  ${\rm Flux(\%)}$\tablenotemark{b}   &  57 \nl \\

  &  Total Flux\tablenotemark{c} & $9.75\times 10^{-7}$   \nl
  &  Total Luminosity\tablenotemark{d} & $2.3 \times 10^{38}$ \nl \cline{1-3}
  &  $\chi^2/\nu$  & 491.7/510   \nl
\enddata
\tablenotetext{a} {The Cosmic X-ray Background (CXB)
component (not shown in the table) is frozen in the model with a flux of
$6.2 \times 10^{-8} \, {\rm ergs \, cm^{-2} \, s^{-1} \,
sr^{-1}}$ in the $2-10$~keV band.}
\tablenotetext{b} {Percentage of the flux of this component
to the total flux.}
\tablenotetext{c} {Total unabsorbed flux in the indicated band in
${\rm ergs \, cm^{-2} \, s^{-1} \, sr^{-1}}$.}
\tablenotetext{d} {Luminosity in ${\rm ergs \,
s^{-1}}$ was calculated assuming
an emitting volume of $10^{66} \, {\rm cm^{-3}}$ and an average depth of
emission of $\sim 17$~kpc.}

\end{deluxetable}
}


\clearpage

\begin{thebibliography}{}

\bibitem[and]{and} 
Anders, E., \& Grevesse, N. 1989, Geochim. Cosmochim. Acta, 53, 197


\bibitem[berger]{berger}
Berger, M. J., \& Seltzer, S. M. 1982, National Bureau of Standards Rep.:
NBSIR 82-2550

\bibitem[bleach]{bleach}
Bleach, R. D., Boldt, E. A., Holt, S. S., Schwartz, D. A., \& Serlemitsos,
P. J. 1972, ApJ, 174, L101

\bibitem[boggs]{boggs}
Boggs, S. E., Lin, R. P., Slassi-Sennou, S., Coburn, W., \& Pelling, R. M.
1999, astro-ph/9908170

\bibitem[dogiel]{dogiel}
Dogiel, V. A., Ichimura, A., Inoue, H., \& Masai, K. 1998, PASJ, 50, 567 

\bibitem[harris]{harris}
Harris, M. J., et al. 1997, in Proc. Fourth Compton Symp., (New York: AIP),
2, 1079

\bibitem[Kaneda]{Kaneda}
Kaneda, H. 1997, Ph.D. Dissertation, University of Tokyo  

\bibitem[kan]{kan}
Kaneda, H., Makishima, K., Yamauchi, S., Koyama, K., Matsuzaki, K., \&
Yamasaki, N. Y. 1997, ApJ, 491, 638 (K97)

\bibitem[kaas]{kaas}
Kaastra, J. S., 1992, An X-ray Spectral Code for Optically Thin Plasmas
(Internal SRON-Leiden Report, updated version 2.0)

\bibitem[koch]{koch} 
Koch, H. W., \& Motz, J. W. 1959, Rev. Mod. Phys., 31, 920

\bibitem[koyama]{koyama}
Koyama, K., Makishima, K., Tanaka, Y., \& Tsunemi, H. 1986, PASJ, 38, 121

\bibitem[krau]{krau} 
Krause, M. O. 1979, J. Phys. Chem. Ref. Data, 8, 307

\bibitem[long]{long}
Long, X., Liu, M., Ho, F., \& Peng, X. 1990, 
Atomic Data and Nucl. Data Tables, 45, 353

\bibitem[mak]{mak}
Makishima, K. et al. 1996, PASJ, 48, 171 

\bibitem[mewe]{mewe}
Mewe, R., Gronenschild, E. H. B. M., \& van den Oord, G. H. J. 1985,
A\&AS, 62, 197

\bibitem[mewe2]{mewe2}
Mewe, R., Lemen, J. R., \& van den Oord, G. H. J., 1986, A\&AS, 65, 511
 
\bibitem[Ottman]{ottman}
Ottmann, R., \& Schmitt, J. H. M. M. 1992, A\&A, 256, 421 

\bibitem[pari]{pari}
Parizot, E. M. G., Cass\'e, M., \& Vangioni-Flam, E. 1997, A\&A, 328, 107

\bibitem[Patterson]{patt}
Patterson, J. 1984, ApJS, 54, 443  

\bibitem[pohl]{pohl}
Pohl, M. 1998, A\&A, 339, 587

\bibitem[quarles]{quarles}
Quarles, C. A. 1976, Phys. Rev. A, 13, 1278

\bibitem[ram2]{ram2}
Ramaty, R., Kozlovsky, B., Lingenfelter, R. E., \& Reeves, H. 1997, 
ApJ, 488, 730 

\bibitem[ram3]{ram3}
Ramaty, R., \& Lingenfelter, R. E. 1998, Proceedings of the 
Meudon Symposium, Galaxy Evolution, Kluwer Academic, in press 

\bibitem[salem]{salem} 
Salem, S. I., Panossian, S. L., \& Krause, R. A. 
1974, Atomic Data and Nucl. Data Tables, 14, 91

\bibitem[sch]{sch}
Schlickeiser, R. 1997, A\&A, 319, L5 

\bibitem[skibo]{skibo}
Skibo, J. G., \& Ramaty, R. 1993, A\&AS, 97, 145

\bibitem[skibo2]{skibo2}
Skibo, J. G., Ramaty, R., \& Purcell, W. R. 1996, A\&AS, 120, 403

\bibitem[tanaka]{tanaka} 
Tanaka, Y., Miyaji, \& Hasinger 1999, in the proceedings of the 4th ASCA
Symposium, in press  

\bibitem[tanuma]{tanuma}
Tanuma, S., Yokoyama, T., Kudoh, T., Matsumoto, R., Shibata, K., \& 
Makishima, K. 1999, PASJ, 51, 161

\bibitem[tati]{tati}
Tatischeff, V., Ramaty, R., \& Kozlovsky, B. 1998, ApJ, 504, 874 

\bibitem[tat]{tat}
Tatischeff, V., Ramaty, R., \& Valinia, A., 1999, in "LiBeB, Cosmic Rays and
Gamma-Ray Line Astronomy", ASP Conference Series, eds. R. Ramaty, E. Vangioni-Flam, 
M. Cass\'e and K. Olive, 226

\bibitem[Townes]{townes}
Townes, C. H. 1989, in IAU Symp. 136, The Center of the Galaxy,
ed. M. Morris (Dordrecht,: Kluwer), 1

\bibitem[val]{val}
Valinia, A., \& Marshall, F. E. 1998, ApJ, 505, 134 

\bibitem[val2]{val2}
Valinia, A., Kinzer, R. L., \& Marshall, F. E. 2000, ApJ, 534, 277 (VKM00)

\bibitem[wd]{wd}
Wdowczyk, J., \& Wolfendale, A. W. 1989, Annu. Rev. Nucl. Part. Sci., 39, 43

\bibitem[Yamasaki et al]{Yamasaki}
Yamasaki N. Y. et al. 1997, ApJ, 481, 821  

\end{thebibliography}
\end{document}